\documentclass[
aps,
pra,
reprint,
twoside,
showkeys,
nofootinbib,
floatfix
]{revtex4-2}    

\usepackage[T1]{fontenc}
\usepackage[utf8]{inputenc}  
\usepackage[english]{babel}
\usepackage{amsmath,amssymb,amsfonts}
\usepackage{mathtools}
\usepackage[dvipsnames]{xcolor}
\usepackage{physics}
\usepackage{graphicx}
\usepackage[normalem]{ulem}
\usepackage{booktabs}
\usepackage{multirow}
\usepackage{csquotes}
\usepackage{afterpage}
\usepackage{placeins}
\usepackage{float}

\newcommand*{\mH}{\emph{m}HAVOK }

\usepackage[
pdftex,
pdftitle={Spectral analysis of the Koopman operator as a framework for recovering Hamiltonian parameters in open quantum systems},
pdfauthor={Jorge E. Pérez-García, Carlos Colchero and Julio C. Gutiérrez-Vega},
colorlinks=true,
linkcolor=blue,
citecolor=blue,
urlcolor=black
]{hyperref}

\begin{document}

\title{Spectral analysis of the Koopman operator as a framework for \\ recovering Hamiltonian parameters in open quantum systems}

\author{Jorge E. Pérez-García}
\author{Carlos Colchero}
\author{Julio C. Gutiérrez-Vega}
\email[]{juliocesar@tec.mx}

\affiliation{
Photonics and Mathematical Optics Group,\texorpdfstring{\\}{ }
Tecnológico de Monterrey, Monterrey, 64849, México.
}

\begin{abstract}
Accurate identification of Hamiltonian parameters is essential for modeling and controlling open quantum systems. In this work, we demonstrate that the multichannel Hankel alternative view of Koopman (\emph{m}HAVOK) algorithm is a robust and reliable spectral data-driven method for retrieving Hamiltonian parameters from the evolution of first-moment observables in open quantum systems. 
The method relies on the discrete spectrum of the Koopman operator to obtain these parameters, which are computed using the \mH algorithm; a theoretical justification for this approach is presented. The method is tested on noiseless quadratures of an open two-dimensional quantum harmonic oscillator and shown to retrieve oscillation frequencies, damping rates, nonlinear Kerr shifts, the qubit-photon coupling strength of a Jaynes-Cummings interaction, and the modulated frequency of a time-dependent Hamiltonian. The majority of the recovered parameters remained within 5\% of their actual values. Compared with Fourier and matrix-pencil estimators, our approach yields lower errors for dynamics with strong dissipation. Overall, these findings suggest that Koopman operator theory provides a practical framework for studying quantum dynamical systems.

\end{abstract}

\date{December 3, 2025}

\maketitle

\newpage
\raggedbottom

\noindent Please cite this paper as:  Jorge E. Pérez-García, Carlos Colchero and J. C. Guti\'errez-Vega, \textit{Spectral analysis of the Koopman operator as a framework for recovering Hamiltonian parameters in open quantum systems}, Phys. Rev. A \textbf{113}, 042222 (2026).
\href{https://doi.org/10.1103/v5zg-gybr}{https://doi.org/10.1103/v5zg-gybr}

\section{Introduction} \label{sec:introduction} 

In recent years, the understanding, characterization, and control of open quantum systems have become relevant topics in circuit and cavity quantum electrodynamics, among many other fields \cite{652674141, PhysRevA.102.012223, RevModPhys.93.025005}. Having an accurate knowledge of oscillation frequencies, dissipation rates, and nonlinear interaction strengths is essential for the experimental calibration of quantum devices and for validating master-equation models. All these topics have also attracted considerable interest in the context of quantum computing \cite{PhysRevApplied.17.054018, PRXQuantum.3.010320}.

In general, parameters of interest are obtained from experimental data using frequency-domain spectroscopic algorithms or nonlinear curve-fitting methods, which often require long acquisition times \cite{PhysRevA.106.022613}. Techniques such as the Fourier, Prony, and matrix-pencil methods have also been used to extract frequencies and decay rates. However, their precision tends to decrease under strong damping conditions \cite{matrixpencil, matrixpencil2}. 
More recently, machine learning and system identification methods have been proposed, but these algorithms often operate as black boxes, lacking physical interpretability and usually relying on large training datasets \cite{626476209, PhysRevResearch.4.043002}. Consequently, there is an increasing interest in fast, data-driven techniques that can directly infer Hamiltonian parameters from experimental data.

In this paper, we demonstrate that the novel multichannel Hankel alternative view of Koopman (\emph{m}HAVOK) algorithm \cite{mHAVOK} is a robust and reliable spectral data-driven method for retrieving Hamiltonian parameters from the evolution of first-moment observables in open quantum systems. Such a technique relies on analyzing the spectrum of the Koopman operator and improves the original HAVOK algorithm \cite{Brunton_Havok_Nature}. A detailed description of its functionality and performance can be found in Ref.~\cite{mHAVOK}.

With this considered, the main contributions of this work are summarized as follows: 

(a) First, we demonstrate the theoretical connection between the Koopman operator spectrum and the \mH model. To the best of our knowledge, this relationship has not been formally established in the literature. However, this step is essential, as it justifies why the discrete spectrum of the Koopman operator is retrieved by one of the dynamical matrices generated by the \mH method. 
    
(b) Second, we show that this spectrum accurately recovers the Hamiltonian parameters of open quantum systems. To test the method, we simulated the dynamics of an open two-dimensional Quantum Harmonic Oscillator (2D QHO) with multiple nonlinear effects and temporal perturbations to gather relevant data. Specifically, the algorithm was validated on the expected values of noiseless quadratures, retrieving oscillation frequencies, damping rates, nonlinear Kerr shifts, qubit-photon coupling strengths in a Jaynes-Cummings interaction, and the modulated frequency of a time-dependent Hamiltonian. The majority of the recovered parameters remained within 5\% of their actual values. Compared with Fourier and matrix-pencil estimators, our approach yielded lower errors for dynamics with strong dissipation.
    
(c) Third, we contribute to reducing the demand for fast, data-driven techniques capable of inferring Hamiltonian parameters directly from experimental data. Since closed-form solutions for the master equations governing the evolution of quantum systems are often unavailable in real-world experimental settings, our method offers a data-driven, spectral-based approach to recover system parameters. For example, quadrature measurements could be obtained by homodyne detection \cite{homodyne_detection} and subsequently analyzed using our framework, with the appropriate denoising filters applied.

The paper is organized as follows: 
Section~\ref{sec:connection_Koopman_mHAVOK} provides a mathematical justification on why the spectral analysis performed on \mH is equivalent to studying the spectrum of the Koopman operator. 
In Sect.~\ref{sec:theory_open_quantum_systems}, we briefly review some basic concepts of open quantum systems and the Lindblad master equation to establish notation and provide necessary formulas.
The numerical results of the simulations and the operation of the \mH method are included in Sect.~\ref{sec:results} for the case of both time-independent and time-dependent Hamiltonians. 
Section \ref{sec:conclusions} summarizes the main contributions and limitations of this work. Finally, to keep this work relatively self-contained, we provide a brief overview of the \mH method in Appendix~\ref{sec:appendixA}.


\section{Demonstration of the Koopman - \mH connection} \label{sec:connection_Koopman_mHAVOK}

We begin by demonstrating the connection between the spectrum of the Koopman operator and the \mH algorithm. This connection is crucial, as it provides the theoretical foundation for applying \mH to the open quantum systems we are analyzing.

\subsection{Concepts of Koopman operator theory}

Consider a smooth dynamical system of the form $\dot{\mathbf{x}}=\mathbf{F}(\mathbf{x})$  evolving on a state space $\mathcal{M}$. Here, $\mathbf{F}$ represents the vector field, $\mathbf{x}$ the state vector, and $\mathcal{M}$ is assumed to be a smooth Riemannian manifold. 

 We now define a smooth function of the state vector $g:\mathcal{M}\rightarrow \mathbb{C}$, which will acquire a central role throughout this work. In the literature of classic dynamical systems, $g=g(\mathbf{x})$ is referred to as an observable, not to be confused with the observables of quantum mechanics. As an example, in a classic system, say the double pendulum, the state vector will consist of position and momentum, and a possible observable could be the energy, since it is a function of those state variables. 

 In order to emphasize the distinction between observables of quantum mechanics and observables in the classic sense, i.e., functions of the state vector, the latter ones will be referred to as K-observables. With this considered, the Koopman semigroup is a linear, infinite-dimensional family of operators that advances in time  K-observables. Let $g\in \mathcal{H}$ represent a  K-observable that lives in a Hilbert space $\mathcal{H}$. The Koopman semigroup $(U^t)_{t\ge0}$ acts on ${g}$ as follows:
\begin{align}
    (U^t g)(\textbf{x})=g\circ\Phi^t(\textbf{x})\label{eq:1},
\end{align}
where $\circ$ represents the composition operator and $\Phi^t$ is the flow map. For a fixed sampling period $\Delta t>0$, we define the discrete Koopman operator as $U^{\Delta t}:=U$. Consequently, along a given 
trajectory $\textbf{x}_k=\Phi^{k\Delta t}(\textbf{x}_0)$, $\textbf{x}_0\in \mathcal{M}$ representing the initial condition, we have
\begin{align}
g(\textbf{x}_k)&=(U^k{g})(\textbf{x}_0),\label{eq:con1}\\
    {g}(\textbf{x}_{k+1})&=(U{g})(\textbf{x}_k).\label{eq:bn_gg}
\end{align}

\subsection{Connection of Koopman theory with delay embeddings}

Suppose that we measure $q$  K-observables $g_j\in\mathcal{H}$, with $j=\{0,1,\cdots,q-1\}$, and collect them into a vector-valued  K-observable $\mathbf{g}:\mathcal{M}\rightarrow \mathbb C^q$ of the form
\begin{align}
    \mathbf{g}(\mathbf{x}) := \big[g_0(\mathbf{x}),\dots,g_{q-1}(\mathbf{x})\big]^T.
\end{align}
We define a delay-embedded  K-observable $\mathbf{h}:\mathcal{M}\rightarrow\mathbb{C}^{mq}$ as
\begin{equation}
    \mathbf{h}(\mathbf{x})
    :=
    \begin{bmatrix}
        \mathbf{g}(\textbf{x}) \\
        \mathbf{g}(\Phi^{\Delta t}(\textbf{x})) \\
        \vdots \\
        \mathbf{g}(\Phi^{(m-1)\Delta t}(\textbf{x}))
    \end{bmatrix}
    =
    \begin{bmatrix}
        \mathbf{g}(\textbf{x}) \\
        (U\mathbf{g})(\textbf{x}) \\
        \vdots \\
        (U^{m-1}\mathbf{g})(\textbf{x})
    \end{bmatrix},\label{eq:ref}
\end{equation}
where $m$ is known as the embedding dimension. Assuming $N$ measurements are made along the trajectory, we define
\begin{align}
    \textbf{g}_k:=\mathbf{g}(\mathbf{x}_k) 
    = \mathbf{g}\big(\Phi^{k\Delta t}(\mathbf{x}_0)\big)\label{eq:flow}
\end{align}
as the $k$-th measurement, with $k$ ranging from $0$ to $N-1$. The corresponding delay-embedded measurements are then given by
\begin{equation}
    \mathbf{h}_k:=\left[\mathbf{g}_{k},\ \mathbf{g}_{k+1},\ \dots,\ \mathbf{g}_{k+(m-1)}\right]^T.
    \label{eq:deylemap}
\end{equation}
All $\mathbf{h}_k$ are then stored in a block Hankel matrix,
\begin{align}
    \mathbf{H} = 
    \begin{bmatrix}
        \mathbf{h}_0 \quad  \mathbf{h}_1 \quad\cdots & \mathbf{h}_{N-m}
    \end{bmatrix} .\label{eq:block_hankel_matrix_generalized}
\end{align}
Given that the delay vectors in Eq.~\eqref{eq:deylemap} are evaluations of the delay  K-observable $\mathbf{h}$ in Eq.~\eqref{eq:ref} along the trajectory, we have
\begin{equation}
    \mathbf{h}_k = \mathbf{h}(\mathbf{x}_k) = \mathbf{h}(\Phi^{k\Delta t}(\mathbf{x}_0))
    = (U^k \mathbf{h})(\mathbf{x}_0).
\end{equation}
Consequently, each column of $\mathbf{H}$ corresponds to the consecutive action of the Koopman operator acting on $\mathbf{h}$ and evaluated at the initial condition $\mathbf x_0$,
\begin{align}
    \mathbf{H} = 
\begin{bmatrix}
    \mathbf{h}(\textbf{x}_0) & (U\mathbf{h})(\textbf{x}_0) &\cdots & (U^{N-m}\mathbf{h})(\textbf{x}_0)
\end{bmatrix} .\label{eq:koopman_blockHankelmatrix}
\end{align}

The columns of \textbf{H} in Eq.~\eqref{eq:koopman_blockHankelmatrix} form a Krylov sequence $\mathcal{F}_{N-m}=\{\textbf{h},U\textbf{h}, \cdots, U^{N-m}\textbf{h}\}\subset \mathcal{H}$ whose columns span a corresponding Krylov subspace \cite{doi:10.1137/17M1125236}. Similarly to the work in Ref.~\cite{doi:10.1137/17M1125236}, we will assume the existence of $\mathcal{K}$, a finite-dimensional subspace of $\mathcal{H}$ that is invariant under the
action of the Koopman operator and whose dimension will be denoted by $d$.  The existence of $\mathcal{K}$ is a strong and mainly theoretical assumption. In practice, the SVD yields the best finite-dimensional approximation even when no exact invariant subspace exists.

Assuming the projection of $\mathbf{h}$ onto $\mathcal{K}$ is cyclic, i.e., repeated applications of the operator generate the whole invariant subspace, we have
\begin{equation}
    \text{span}(\mathcal F_{N-m})=\mathcal{K}
\end{equation}
for $N-m\geq d-1$. 

Next, a singular value decomposition (SVD) is performed on $\mathbf{H}$,
\begin{align}
    \mathbf{H} = \mathbf{U}\mathbf{S}\mathbf{\mathcal{V}}^T,
\end{align}
allowing for the existence of two orthonormal bases for the column and row space of $\mathbf{H}$, each given by the columns of $\mathbf{U}$ and $\mathcal{V}$. If $\sigma_i$ denotes the $i$-th singular value and $v_i[k]$ the $k$-th component of the $i$-th right-singular vector, then each column of the block Hankel matrix can be expressed as
\begin{align}
\mathbf{h}_k=\sum_{i=1}^{qm}\sigma_iv_i[k]\mathbf{u}_i:=\sum_{i=1}^{qm}y_i[k]\mathbf{u}_i,
\end{align}
where 
\begin{equation}
 y_i[k]:=\sigma_iv_i[k]\label{eq:coordinate_vector}   
\end{equation}
is the coordinate vector of $\mathbf{h}_k$ in the basis of $\mathbf{U}$. 

Subsequently, the first $r$ singular vectors are retained in \mH and assumed to be the leading SVD modes that best describe the system dynamics. A fixed value of $r$ is now assumed; its optimal value is discussed in Subsection~\ref{subsec:optimal_cutoff_rank}. Let $\mathcal{H}_P=\text{span}\{ \mathbf{u}_0, \dots, \mathbf{u}_{r-1} \} \subset \mathcal{H}$ denote the finite-dimensional subspace generated by the retained dynamical modes. 

Reference~\cite{doi:10.1137/17M1125236} showed that for ergodic systems and in the long-time limit, the SVD of the Hankel matrix converges to the Proper Orthogonal Decomposition (POD) of the Krylov sequence. In our case, if $r\ge d$, then a subset of $\mathcal{H}_P$ converges to an orthonormal basis of $\mathcal{K}$. Otherwise, if $r<d$ or if no exact finite-dimensional invariant subspaces exist, then $\mathcal{H}_P$ is the best rank-$r$ approximation to the span of the sampled Krylov sequence. In the following, we assume the existence of an $r\geq d$.

\subsection{Projecting observables onto \texorpdfstring{$\mathcal{K}$}{K}}

Having discussed the assumptions and connections between the SVD modes and Krylov invariant subspaces, two orthogonal projection operators are defined on $\mathcal{H}_p$:
\begin{align}
P&:\mathcal{H}_P\rightarrow\mathcal{K},\\
Q&:\mathcal{H}_P\rightarrow\mathcal{H}_Q,
\end{align}
where $\mathcal{K} \oplus \mathcal{H}_Q=\mathcal{H}_P$ and $Q=I-P$, $I$ representing the identity operator. Notice how $P$ is a projection over the SVD-derived dominant modes that span $\mathcal{K}\subset \mathcal{H}_P$. Let $\mathbf{h}_u\in \mathcal{H}_P$ denote the projection of $\textbf{h}$ onto $\mathcal{H}_P$. Hence, $\textbf{h}_u$ is uniquely decomposed as 
\begin{equation}
\mathbf{h}_u=P\mathbf{h}_u+Q\mathbf{h}_u:=\mathbf{v}+\mathbf{u}, 
\end{equation}
with $\mathbf{v}\in \mathcal{K}$ and $\mathbf{u}\in \mathcal{H}_Q$. Considering the discrete-time Koopman operator acting on $\mathcal{H}_P$, its action can be decomposed as follows:
\begin{align}
    U&=IUI=(P+Q)U(P+Q),\\
    &=PUP+PUQ+QUP+QUQ.\label{eq:fourblocks}
\end{align}

Thus, the projection operator $P$ isolates the part of $\mathbf{h}_u$ that lives in $\mathcal{K}$, while $Q$ collects the residual component. The four blocks in Eq.~\eqref{eq:fourblocks} each describe the action of the operator that (i) stays within $\mathcal{K}$, (ii) maps residual dynamics onto the retained space $(\mathcal{H}_Q\rightarrow \mathcal{K})$, (iii) describes how the retained dynamics interact with the orthogonal complement $(\mathcal{K}\rightarrow \mathcal{H}_Q)$, and (iv) evolves entirely within the residual subspace. 

If $QUP=0$, then $PUP$ gives an exact finite-dimensional restriction of the Koopman operator on that subspace. In the general case, however, $QUP$ is different from zero, thereby being relevant to properly reconstruct the system dynamics, even if they are not captured in a Koopman-invariant subspace. 

Interestingly, the above partition closely resembles the Mori-Zwanzig formalism in statistical physics, which decomposes the system dynamics into resolved and unresolved components via projection operators, yielding closed equations of motion \cite{mori_zwanzig}. Indeed, a connection between Koopman theory and such formalism has already been reported in Ref.~\cite{koopman_moriZwanzig}. 

\subsection{Regression-based approximation of the Koopman operator acting on \texorpdfstring{$\mathcal{K}$}{K}}

Having discussed the projection operators, the resolved part after one step is given by 
\begin{align}
    P(U\textbf{h}_u)=PUP \textbf{v} + PUQ \textbf{u}.\label{eq:forced_linear_model_1}
\end{align}

The \mH algorithm constructs a forced linear model on the time series of the leading SVD coordinates, i.e., the first $r$ right-singular vectors, in analogy to Eq.~\eqref{eq:forced_linear_model_1}. The linear part of the model describes the dynamics that evolve under $\mathcal{K}$, while the forcing is assumed to come from its orthogonal projection. 

Recalling the assumption that a subspace of $\mathcal{H}_P$ spans $\mathcal{K}$, the method performs a regression step [Eq.~\eqref{eq:minimization}] to classify the $i$-th main dynamical mode (out of the $r$ ones) as linear or nonlinear. The linear modes are assumed to span $\mathcal{K}$, while the nonlinear modes capture the unresolved dynamics.

Following Eq.~\eqref{eq:coordinate_vector}, if $r_c$ and $r_f$ each denote the index sets of linear and nonlinear components, then the following coordinate vectors are defined:
\begin{align}
    \mathbf{y}_k&:=[y_j[k]: j\in r_c]^T\in \mathbb{C}^{|r_c|},\\
    \mathbf{u}_k&:=[y_j[k]: j\in r_f]^T\in \mathbb{C}^{|r_f|},
\end{align}
with $|\cdot|$ representing the cardinality of the sets. With this considered, the algorithm models the dynamics as
\begin{align}
    \textbf{y}_{k+1}=\mathbf{A} \textbf{y}_k + \mathbf{B} \textbf{u}_k,
\end{align}
where $\mathbf{A}$ and $\mathbf{B}$ are obtained from a least-squares regression. Considering $\mathbf{y}_k$ and $\mathbf{u}_k$ are coordinate vectors of $P\mathbf{h}_u$ and $Q\mathbf{h}_u$ in the $\{\mathbf{u}_j\}$ basis, $\mathbf{A}$ and $\mathbf{B}$ can be thought of as finite-dimensional approximations to the Koopman blocks
\begin{align}
    \mathbf{A} \approx PUP\big|_{\mathcal{K}},
    \qquad
    \mathbf{B} \approx PUQ\big|_{\mathcal{H}_Q}.
\end{align}

In this view, $\mathbf{A}$ approximates the action of $U$ on the invariant subspace spanned by the linear modes, while $\mathbf{B}$ quantifies how the nonlinear components influence the resolved coordinates. 

Moreover, the spectrum of $\mathbf{A}$ approximates the spectrum of the Koopman operator restricted to $\mathcal{K}$. Let
\begin{equation}
    \sigma(\mathbf{A})=\{\lambda \in \mathbb{C}\big|\det (\mathbf{A}-\lambda I)=0\} \label{eq:eigvals_A}
\end{equation}
represent the set of retrieved eigenvalues $\lambda_i$. Although the Koopman operator has both a continuous and a discrete spectrum \cite{Mezic2005}, $\mathbf{A}$ recovers the latter, since finding a finite-dimensional realization of the Koopman operator in a Koopman-invariant subspace is, in general, equivalent to the existence of the pure point spectrum in that subspace \cite{doi:10.1137/17M1125236}.  Furthermore, on the invariant subspace $\mathcal{K}$, the evolution of the projection of any $K$-observable onto $\mathcal{K}$ can be expressed as
\begin{equation}
     g(t)\approx \sum_j c_j e^{\lambda_jt}\phi_j\label{eq:koopman_mode_Decomposition},
\end{equation}
with $c_j\in \mathbb{C}$ and $\phi_j$ representing the Koopman eigenfunction associated with the $j$-th eigenvalue.

It is important to note, however, that the existence of a purely discrete spectrum on a dynamical system is largely a theoretical condition \cite{Haller2024}. In practice, the majority of the systems exhibit both a continuous and discrete spectrum. \mH overcomes such a limitation by allowing the existence of $\mathbf{B}$, a (possibly) rectangular matrix that contains information of nonlinear components. Hence, we hypothesize that such a matrix might be related to the continuous spectrum. Considering that the spectral analysis of $\mathbf{A}$ alone is sufficient for a proper recovery of Hamiltonian parameters, as will be shown later, such a hypothesis is therefore not pursued in the current work.

In summary, by obtaining the eigenvalues of Eq.~\eqref{eq:eigvals_A}, we approximated the pure point spectrum of the Koopman operator and thus established the link between Koopman operator theory and the Hankel matrix from the \mH method. To the best of our knowledge, this theoretical connection has not been presented yet in the literature and thus is an original contribution of this work. In the following, we will use these results to interpret the retrieval of Hamiltonian parameters in open quantum systems with the \mH algorithm.


\section{Preliminary concepts and considerations of the problem} \label{sec:theory_open_quantum_systems}

In this section, we briefly recall some basic concepts of open quantum systems to establish notation and provide necessary formulas.

The dynamics of open quantum systems can be characterized by the Lindblad master equation, mainly in the Born and Markov approximations \cite{Lindblad1976}. Its standard form reads as
\begin{align}
    \frac{d\rho}{dt}=-\frac{i}{\hbar}[H, \rho]+\sum_j\left(L_j\rho L_j^\dagger - \frac{1}{2}\left\{L_jL_j^\dagger, \rho\right\}\right),\label{eq:linbladian}
\end{align}
where $\rho$ is the density matrix, $H$ is the Hamiltonian, $\{\cdot,\cdot\}$ denotes the anti-commutator, $L_j$ are jump operators that model system-environment interactions and $\cdot^\dagger$ represents the complex conjugate. The master equation can be shortly written as 
\begin{equation}
 \dot{\rho}=\mathcal{L}(\rho),\label{eq:t56}
\end{equation}
with the Liouvillian superoperator $\mathcal{L}$ acting linearly on $\rho$. Hence, the solution of Eq.~\eqref{eq:linbladian} reads as follows:
\begin{equation}
    \rho(t)=e^{\mathcal{L}t}\rho(0).\label{eq:29}
\end{equation}
A comprehensive derivation of Eqs.~\eqref{eq:linbladian} -~\eqref{eq:29} can be found in Ref.~\cite{short_intro_lindblad}; for a deeper explanation, we refer the reader to Ref.~\cite{breuer}.

In the context of Koopman theory, the superoperator $e^{\mathcal{L}t}$ can be viewed as the flow map $\Phi^t$ from Eq.~\eqref{eq:1}, but now acting on the state space of density matrices. With this considered, the solution of the Lindbladian can be expressed as 
\begin{equation}
\rho(t)=e^{\mathcal{L}t}\rho(0)=\Phi_t(\rho_0). 
\end{equation}

 Moreover, the solution of the eigenvalue problem $\mathcal{L}(R_j)=\lambda_jR_j$ allows for expressing the density matrix as
\begin{equation}
     \rho(t)=\sum_j \tilde{c}_je^{\lambda_jt}R_j,\label{eq:linbladian2}
\end{equation}
with both $\tilde{c}\land \lambda_j\in \mathbb{C}$. Expressing the density matrix in the basis of the Liouvillian eigenoperators strongly resembles the approximation of a $K$-observable in terms of the Koopman eigenfunctions [Eq.~\eqref{eq:koopman_mode_Decomposition}]. Indeed, as the $K$-observables analyzed in this work are functions of the density matrix, i.e., $g(t)=g((\rho(t))$, 
their time evolution inherits the same exponential rates as in Eq.~\eqref{eq:linbladian2}. Consequently, the real and imaginary parts of the eigenvalues of the matrix $\mathbf{A}$ obtained via mHAVOK, having proved to approximate the discrete spectrum of the Koopman generator in Sec.~\ref{sec:connection_Koopman_mHAVOK}, correspond directly to the physical decay rates and oscillation frequencies given by the exponential rate constants $\lambda_j$ in Eq.~\eqref{eq:linbladian2}. As will be discussed later, this theoretical connection allows these parameters to be obtained via the \mH method.

Throughout this work, we model an open 2D QHO solving Eq.~\eqref{eq:linbladian} under a common environment for both the $x$ and $y$ Cartesian components of the oscillator. The nature of the bath is described as follows:

The emission of photons from the system to the bath, also known as a decay, is given by
\begin{align}
    \mathcal{L}_c=a_x+e^{i\phi}a_y,\label{eq:common_environment}
\end{align}
with $\phi$ being the relative phase between the two system modes and $a_i$ representing the annihilation operator for the \textit{i}-th component. Physically, the bath does not recognize whether the emission came from the $x$ or $y$ component; it only couples both emissions to the linear combination in Eq.~\eqref{eq:common_environment}. On the other hand, absorption of photons from the bath to the system, a process also known as excitation, is given by $\mathcal{L}_c^\dagger$.

Additionally, the bath has a temperature $T>0$ and is modeled as a collection of harmonic oscillators, from which the system can both emit and absorb quanta. The average number of photons in the bath at frequency $\omega_i$ is given by the Bose-Einstein distribution \cite{breuer}
\begin{align}
    \overline{n}_i(\omega_i,t)=\left[\exp\left( \frac{\hbar \omega_i}{k_BT} \right)-1\right]^{-1},
\end{align}
where $i=\{x,y\}$ and $k_B$ is the Boltzmann constant. Taking this into account, the jump operators defined in Eq.~\eqref{eq:linbladian} are given by
\begin{align}
    L_i^\text{decay}&:=\mathcal{L}_c \sqrt{\kappa(\overline{n}_i+1)},\label{eq:jump_operator_decay}\\
    L_i^\text{excitation}&:= \mathcal{L}_c^\dagger\sqrt{\kappa\overline{n}_i},\label{eq:jump_operator_excitation}
\end{align}
with $\kappa$ denoting the damping rate. Note that decay occurs even in the absence of photons, such as in a vacuum. These phenomena can be attributed to quantum fluctuations within the system \cite{breuer}.

\begin{figure*}[htbp]
\centering
\includegraphics[width=\linewidth]{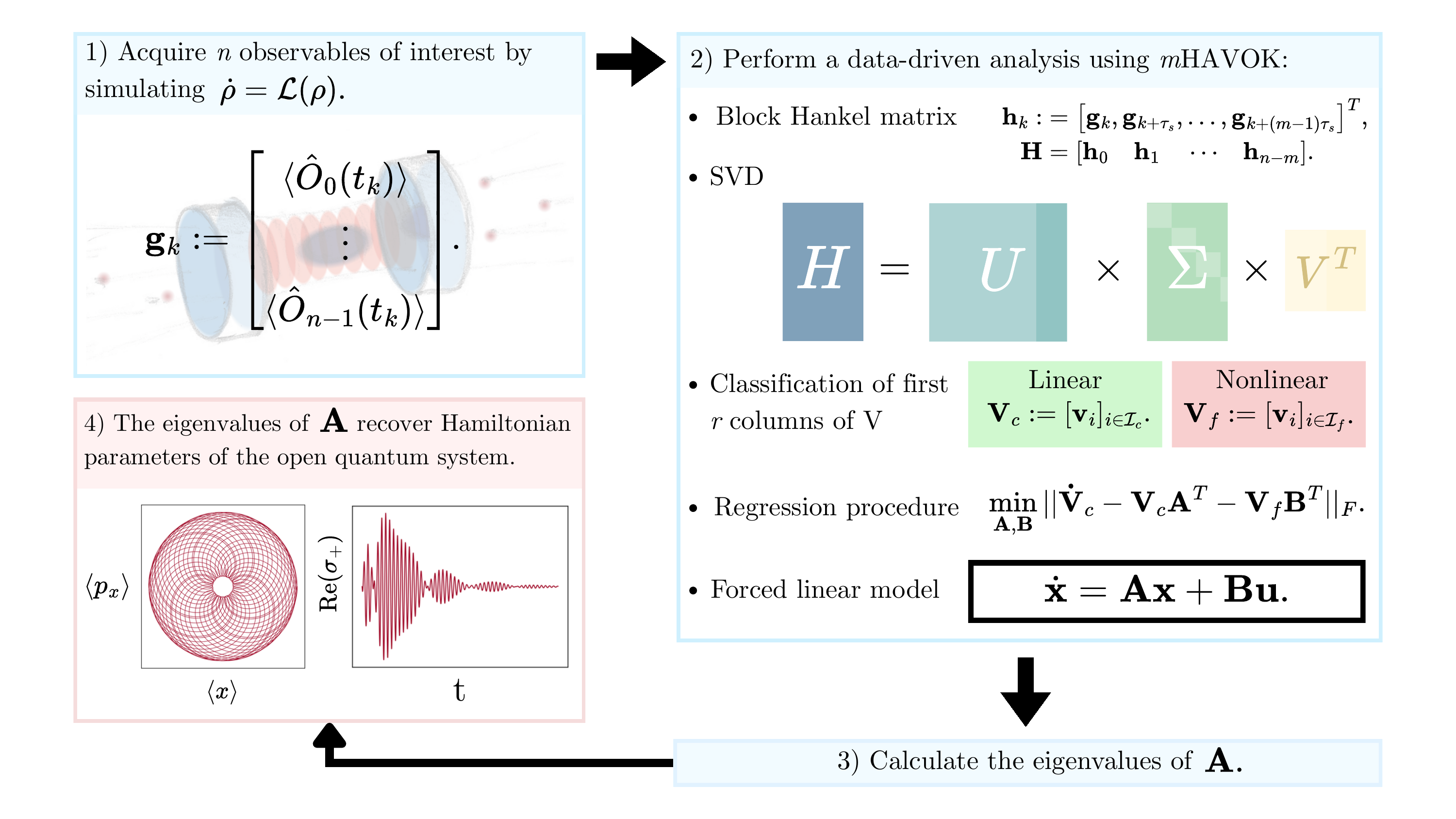}
\caption{Overview of the pipeline employed in this work. (1) The expected values of observables are first obtained by simulating the Lindblad master equation. (2) Such data is then analyzed through the \mH framework, where two dynamical matrices are retrieved.  The eigenvalues are calculated for one of such matrices (3), allowing recovery of the Koopman generator's discrete spectrum and, thus, the Hamiltonian parameters of the open quantum system (4).}
\label{fig:diagram_summarize}
\end{figure*}


\section{Retrieving Hamiltonian parameters}\label{sec:results} 

The spectral analysis performed to retrieve the Hamiltonian parameters using \mH is illustrated in Fig.~\ref{fig:diagram_summarize}. We begin by generating the expected values for $n$ observables of the open quantum system of interest using an accurate numerical solver of the Lindblad master equation \cite{qutip}. This data is then used as input for the \mH algorithm, which allows us to (1) retrieve a forced linear model of the system dynamics and (2) utilize the spectrum of the matrix $\mathbf{A}$ to estimate the desired Hamiltonian parameters for the open quantum system.  It is worth emphasizing that the recovery of such parameters is entirely data-driven, as the algorithm only requires time series of the measured observables and does not rely on analytical solutions of their dynamics.

In this work, we will consider a 2D QHO as our test system, incorporating various time-independent and time-dependent nonlinear effects. The initial state chosen for our simulations is a coherent state with an eigenvalue of $\alpha=1$. The parameters used in these simulations are detailed in Table~\ref{tab:simulation_parameters}. The evolution of the quadrature phase-space vector
\begin{align}
    \mathbf{x}=\begin{bmatrix}
    \expval{x(t)}\\
    \expval{y(t)}\\
    \expval{p_x(t)}\\
    \expval{p_y(t)}
    \end{bmatrix}\label{eq:quadrature_vector}
\end{align} is obtained by solving the Lindblad master equation Eq.~\eqref{eq:linbladian} with the QuTiP solver \cite{qutip}. This data is then provided as input to the \mH algorithm for all the studied systems, except for a Jaynes-Cummings interaction (see \ref{subsec:JC_results}).
The user-defined threshold used in the algorithm for discerning linear from nonlinear components was set to $\tau=0.95$;  the purpose of this parameter is further explained in Appendix~\ref{sec:appendixA}. Additionally, the embedding dimension was set to $m=100$, consistent with previous research on the HAVOK method \cite{mHAVOK}. This parameter is also part of the \mH input configuration and is further elaborated in Appendix~\ref{sec:appendixA}.

\begin{table}[t]
\caption{Simulation parameters employed in this work.}\label{tab:simulation_parameters}
\centering
\rule{\linewidth}{0.5pt}
\begin{tabular}{@{}c c c c@{}}
    Variable & \hspace{7em} &  & Value \\
\end{tabular}
\vspace{0.3em}
\rule{\linewidth}{0.4pt} 
\begin{tabular}{@{}llc@{}}
    \multicolumn{2}{l}{Simulation time} & $t_f = 50$ \\
    \multicolumn{2}{l}{Time step} & $\Delta t = 0.01$ \\
    \multicolumn{2}{l}{\multirow{2}{*}{Oscillation frequencies}} & $\omega_x = 2\pi$ \\
    \multicolumn{2}{l}{} & $\omega_y = \pi$ \\
    \multicolumn{2}{l}{Environment temperature} & $T = 2.0$ \\
    \multicolumn{2}{l}{Damping rate} & $\kappa = 0.1$ \\
    \multicolumn{2}{l}{\multirow{2}{*}{Fock-state truncations}} & $n_x = 10$ \\
    \multicolumn{2}{l}{} & $n_y = 10$ \\
\end{tabular}
\rule{\linewidth}{0.5pt}
\end{table}

In the following, a hat over a variable indicates that the corresponding value was numerically recovered from the spectrum of $\mathbf{A}$. Conversely, if the variable lacks a hat, it represents the original parameter used to solve the Lindbladian.  With this considered, the percent error between the two quantities will be denoted as $e_{\lambda\hat{\lambda}}$ and calculated as
\begin{equation}
     e_{\lambda \hat{\lambda}}= \frac{|\hat{\lambda}-\lambda|}{\lambda}\cdot100,
\end{equation}
with $|\cdot|$ denoting the absolute value. In general, the acceptable value of the obtained percent error is application dependent. In high-fidelity quantum control settings (e.g., qubit calibration for high-fidelity gates), sub-percent precision is typically required, as coherent frequency or coupling mismatches have a direct impact on gate infidelities~\cite{Chow2008RB,Werninghaus2021PRXQ}. In contrast, for system identification, model validation, or initial quantum device characterization, percent-level accuracy is often sufficient for capturing the main dynamical parameters \cite{PhysRevLett.109.240504}, especially when finite coherence times limit spectral resolution.

In the following, multiple examples will demonstrate how \mH enables accurate parameter recovery, even in the presence of nonlinearities and time-dependent Hamiltonians. The percent error obtained will be compared with that from traditional methods for recovering oscillation frequencies, such as the Fast Fourier Transform (FFT) and the generalized pencil-of-function (matrix pencil) method \cite{matrixpencil}. Specifically, the matrix pencil method is an algorithm that effectively extracts damping factors and oscillation frequencies from an input signal, assuming that a sum of complex exponentials provides a suitable basis for it \cite{matrixpencil2}.

\subsection{Two-dimensional quantum harmonic oscillator}\label{subsec_results_2D_QHO}

The Hamiltonian of a 2D QHO is given by 
\begin{align}
    H_\text{0}=\hbar\omega_x{a}_x{a}^\dagger_x+\hbar\omega_y {a}_y{a}^\dagger_y, \label{eq:hamiltonian_field}
\end{align}
where $\omega_i$ are the oscillation frequencies and $a_i^\dagger$ denote the creation operators. For notational convenience, the zero of the energy has been shifted upwards by $\hbar(\omega_x+\omega_y)/2$, merely by redefining the reference of the energy. Throughout this work, the ladder operators and, hence, the quadratures are expressed in dimensionless units. 

Considering that the Hamiltonian in Eq.~\eqref{eq:hamiltonian_field} is quadratic and the jump operators in Eqs.~\eqref{eq:jump_operator_decay} and~\eqref{eq:jump_operator_excitation} are linear, then the Heisenberg equations of motion for the first moments evolve linearly and have closed-form expressions \cite{breuer}. Consequently, the evolution of the quadrature vector Eq.~\eqref{eq:quadrature_vector}
is given by the following ordinary differential equation (ODE):
\begin{align}
    \dot{\mathbf{x}}= \mathbf{S}~\mathbf{x},\label{eq:linear_ode}
\end{align}
where
\begin{align}
    \mathbf{S} &=
    \begin{bmatrix}
  -\kappa/2 & \omega_x & 0 & 0 \\
  -\omega_x & -\kappa/2 & 0 & 0 \\
  0 & 0 & -\kappa/2 & \omega_y \\
  0 & 0 & -\omega_y & -\kappa/2 \\
    \end{bmatrix}.
\end{align}

\begin{figure}[t]
    \centering
    \includegraphics[width=\linewidth]{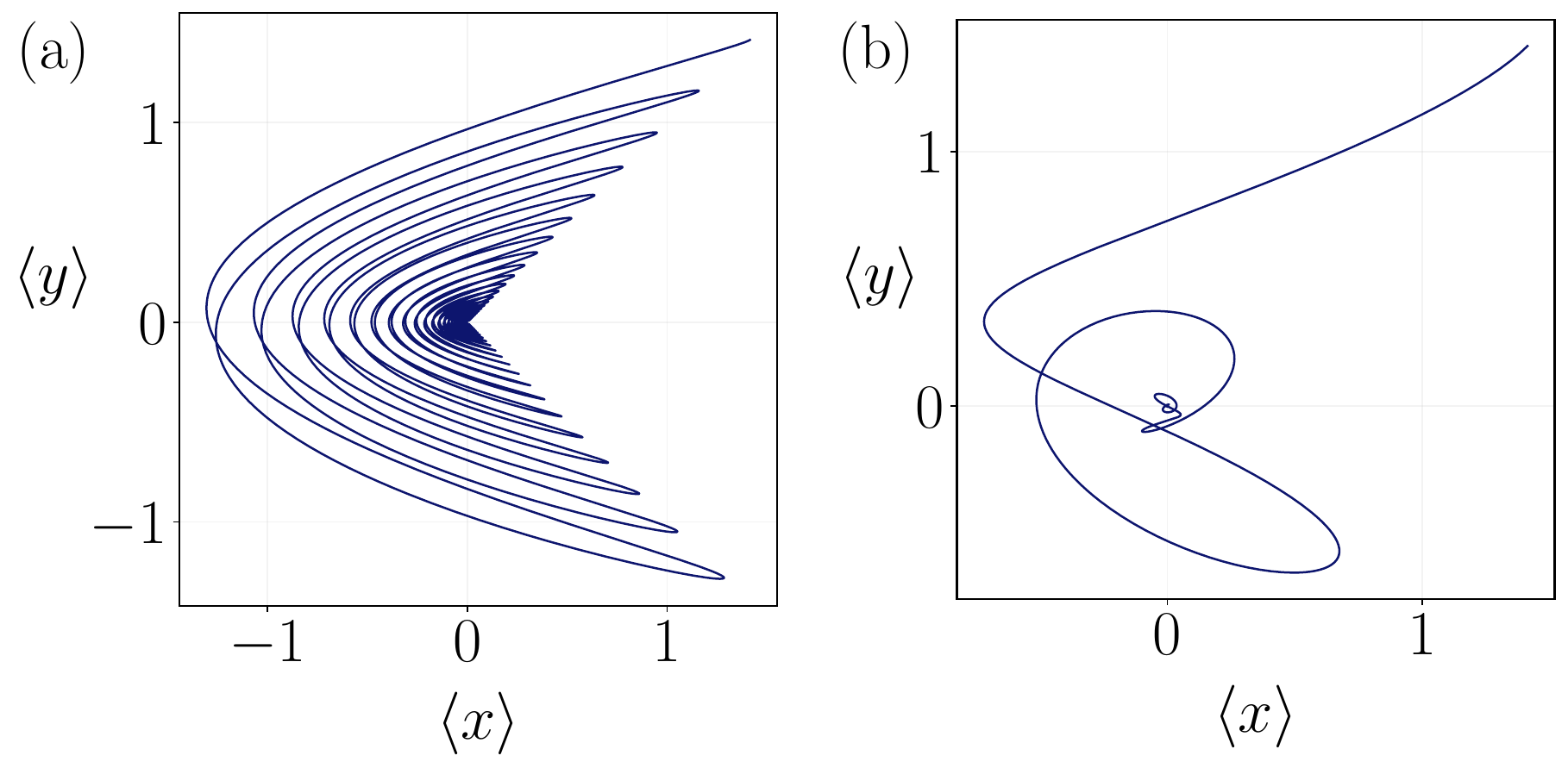}
    \caption{\mH  phase-space reconstruction of an open 2D QHO for the $x$ and $y$ components using a damping rate of (a) $\kappa=0.1$ and (b) $\kappa=1.0$. No nonlinearities were introduced.}
    \label{fig:firstQHO_simulation}
\end{figure}

Note that $\mathbf{S}$ depends on the oscillation frequencies and damping coefficients, but not on the bath occupations $\overline{n}_i$. In Fig.~\ref{fig:firstQHO_simulation}, we illustrate the system evolution for two different values of the damping rate $\kappa$ and data taken from Table \ref{tab:simulation_parameters}. The subplot~\ref{fig:firstQHO_simulation}\textcolor{blue}{(a)} shows the trajectory of the expected value of the particle in the $(x,y)$ plane, where we see that the $x$-mode oscillates twice as fast as the $y$-mode, in agreement with the simulated oscillation frequencies. Furthermore, the system follows a parabolic Lissajous curve, with the amplitudes decaying and collapsing towards the origin due to the open setup. In contrast, Fig.~\ref{fig:firstQHO_simulation}\textcolor{blue}{(b)} shows a substantially faster collapse towards the origin resulting from an order-of-magnitude increase in the damping rate. Consequently, no parabolic trajectories are identified.

Regarding the \mH model, the optimal cutoff rank $r_\text{opt}$ was found to be $r=4$, which coincides with the number of input observables. This correspondence is not a general feature but rather a consequence of the system dynamics: since the input quadrature vector obeys the linear ODE in Eq.~\eqref{eq:linear_ode}, higher-order modes do not contribute significantly to the reconstructed dynamics. 

After determining the optimal cutoff rank, we obtained the retrieved dynamical matrix $\mathbf{A}$. By applying Schur decomposition \cite{horn_johnson}, we found that this matrix takes the form of $\mathbf{S}$ as in Eq.~\eqref{eq:linear_ode}. Furthermore, its four eigenvalues, up to numerical error, appeared in conjugate pairs of the form $-\hat{\kappa}/2 \pm \hat{\omega}_i$. As discussed in Sect.~\ref{sec:connection_Koopman_mHAVOK}, these eigenvalues correspond to those of the Koopman operator restricted to the invariant subspace spanned by our observables, thereby providing a linear description of the system dynamics.

\begin{figure}[t]
    \centering
    \includegraphics[width=0.97\linewidth]{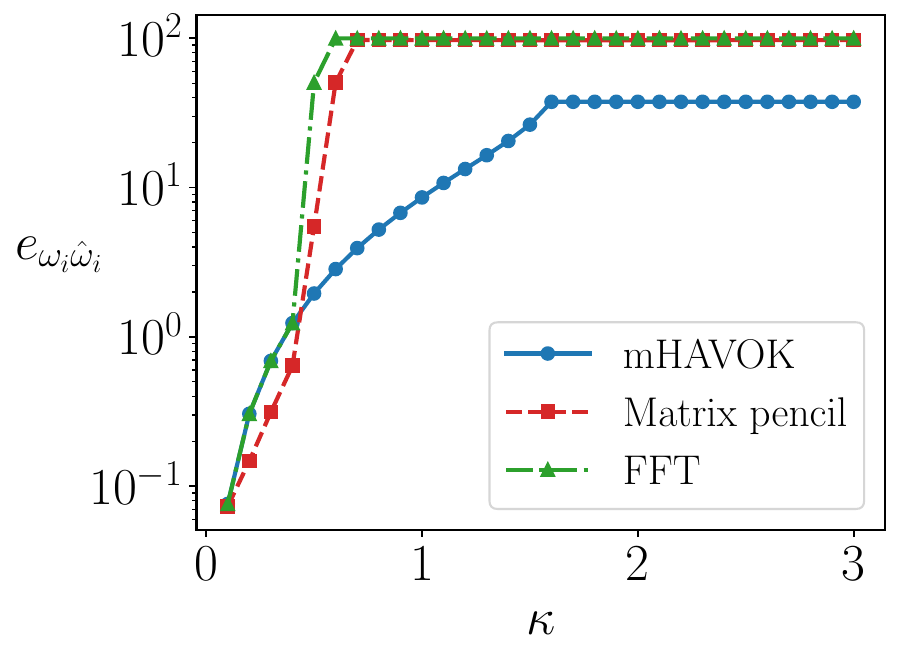}
    \caption{Percent error comparison among three different methods for determining the oscillation frequencies $\hat{\omega}_i$ of the 2D QHO. \mH exhibits a slower increase in percent error compared to existing alternatives. }
    \label{fig:percenterror_comparison}
\end{figure}

Once the spectrum of $\mathbf{A}$ was obtained, the percent error between the retrieved Hamiltonian parameters and their true values was calculated. For the case in Fig.~\ref{fig:firstQHO_simulation}\textcolor{blue}{(a)} where $\kappa=0.1$, the average percent error between $\omega_i$ and $\hat{\omega}_i$ was $e_{\omega_i\hat{\omega}_i}=0.08\%$, whereas $e_{\kappa\hat{\kappa}}=7.2\times10^{-4} \%$. On the other hand, when increasing 10 times the value of $\kappa$ in the simulation setup [Fig.~\ref{fig:firstQHO_simulation}\textcolor{blue}{(b)}], a notable increase is observed in the oscillation-frequency error, with $e_{\omega_i\hat{\omega_i}}=8.58\%$. Interestingly, the error for the damping rate decreased almost four times once we increased its value, with $e_{\kappa\hat{\kappa}}=1.8\times10^{-4}\%$. Figure~\ref{fig:percenterror_comparison} compares the percent error associated with three different methods for retrieving the oscillation frequencies as the damping rate increases. Although the matrix-pencil scheme is more effective for weak damping rates ($\kappa<0.5$), \mH significantly improves recovery at stronger damping rates.

\subsection{Kerr nonlinearities}\label{subsec:Kerr_results}

In classical optics, the Kerr effect appears in a nonlinear medium whose refractive index depends on the intensity of the electric field \cite{walls2008quantum}. In quantum optics, this effect shifts the oscillation frequencies by an amount proportional to the number of photons present at a given time. The Hamiltonian that describes this phenomenon is given by \cite{Kerr_Nonlinearity}:
\begin{equation}
    H_\text{Kerr} = \hbar \left( \frac{\chi_x}{2} a_x^{\dagger 2} a_x^2 + \frac{\chi_y}{2} a_y^{\dagger 2} a_y^2 + \chi_{xy} n_x n_y \right),  \label{eq:kerrHamiltonian}
\end{equation}
with $n_i=a_ia^\dagger_i$ denoting the number operator and $\chi_i$ representing Kerr strength coefficients. The resulting energy spectrum forms an anharmonic ladder \cite{Carmichael2008}, with the transition frequency between successive Fock states following $\omega_i+\chi_in_i$, with $n_i$ being a natural number. 

The Hamiltonian of the tested system is then
\begin{equation}
    H = H_0+H_\text{Kerr},
\end{equation}
where $H_0$ is given by Eq.~\eqref{eq:hamiltonian_field}. 
Considering that Eq.~\eqref{eq:kerrHamiltonian} contains quartic terms like $a_i^{\dagger2}a_i^2$, the Heisenberg equation of motion for any ladder operator (and hence any quadrature) depends on higher-order moments such as $\expval{a^\dagger aa}$. This dependence prevents the closure of the moment hierarchy, making it impossible to derive a closed expression for their evolution \cite{Carmichael2008}.

Truncating the Heisenberg equations to first-order moments yields a time-dependent shift in the oscillation frequency proportional to the number operator, capturing the intensity-dependent phase rotation characteristic of the Kerr effect. In the steady state, these effective oscillation frequencies take the form
\begin{align}
    \Omega_x^\text{eff} &= \omega_x+\chi_x(\langle n_x \rangle_{ss}-1)+\chi_{xy}\langle n_y \rangle_{ss},\label{eq:effective_frequencies_x}\\
    \Omega_y^\text{eff} &= \omega_y+\chi_y(\langle n_y \rangle_{ss}-1)+\chi_{xy}\langle n_x \rangle_{ss}, \label{eq:effective_frequencies_y}
\end{align}
with $\langle n_i \rangle_{ss}$ representing the mean value of the number operator of the \textit{i}-th mode in the steady state. 

\begin{figure}[t]
    \centering
    \includegraphics[width=\linewidth]{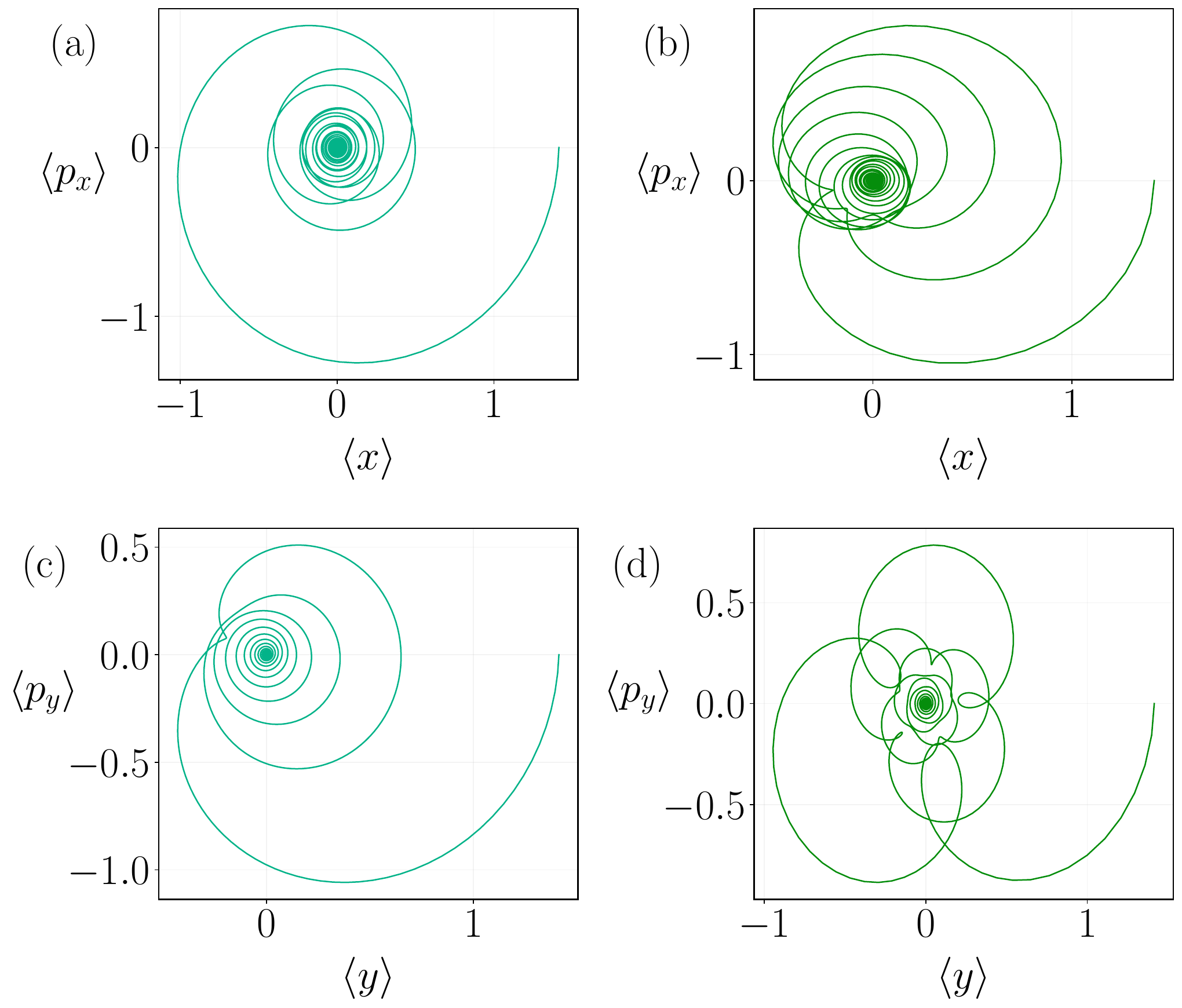}
    \caption{Phase-space reconstruction considering (a-c) $\chi_x,\chi_y=2,3$ and (b-d) $\chi_x=\chi_y=5$. In both cases, the framework successfully recovered Kerr strengths with few numerical errors. No cross-coupling terms were included. }
    \label{fig:chi1}
\end{figure}

Kerr nonlinearities were introduced into the simulations, with strength parameters set to $\chi_x=2$ and $\chi_y=3$, while keeping a damping rate $\kappa=0.1$ and no cross term $\chi_{xy}$. The optimal cutoff rank increased to $r_\text{opt}=20$, five times more than in the previous case. Taking into account that the value of $r_\text{opt}$ determines the number of main dynamical modes identified, four of those 20 were classified as nonlinear. Their appearance can be regarded as a consequence of the lack of a closed-form expression for the evolution of the quadratures, as previously discussed. Although this represents a theoretical limitation, the algorithm can successfully reconstruct the system dynamics using the forced linear model built on the quadratures. Figures~\ref{fig:chi1}\textcolor{blue}{(a)} and~\ref{fig:chi1}\textcolor{blue}{(c)} illustrate the reconstructed phase space, showing a spiral trajectory that collapses smoothly towards the origin, with each turn occurring at a slightly different instantaneous frequency, consistent with the presence of a Kerr nonlinearity.

Regarding the retrieval of Hamiltonian parameters, the eigenvalues of $\mathbf{A}$ appeared in complex-conjugate pairs, similar to the previous case. Their real parts, associated with damping rates, did not exhibit any identifiable pattern. In contrast, the imaginary part followed the relation $\pm (\hat{\omega}_i + n\hat{\chi}_i)$, where $n\in\mathbb{N}$. Hence, \mH was capable of recovering Kerr nonlinearity coefficients alongside the oscillation frequencies. Such a pattern successfully reproduces the characteristic frequency ladder of Kerr-type nonlinear oscillators.

To retrieve the coefficients $\hat{\chi}_i$, the twenty eigenvalues of $\mathbf{A}$ were first sorted by magnitude. For instance, the imaginary parts of the first eight eigenvalues were $\pm\hat{\omega}_y, \pm(\hat{\omega}_y+\hat{\chi}_y), \pm\hat{\omega}_x, \ \text{and}\pm(\hat{\omega}_x+\hat{\chi}_x)$. Considering $\omega_y+\chi_y<\omega_x$, the method ordered the imaginary parts starting from the ones associated with the $y$-mode. Based on this ordering, the Kerr coefficients were estimated by subtracting the absolute values of the imaginary parts of the third and first eigenvalues to obtain $\chi_y$, and similarly, from the seventh and fifth eigenvalues to obtain $\chi_x$. By applying this procedure, the obtained average percent error was $e_{\chi_i\hat{\chi}_i}=0.04\%$. Additionally, the associated error for the oscillation frequencies remained remarkably low, with $e_{\omega_i\hat{\omega}_i}=0.02\%$. 

\begin{figure}[t]
    \centering
    \includegraphics[width=\linewidth]{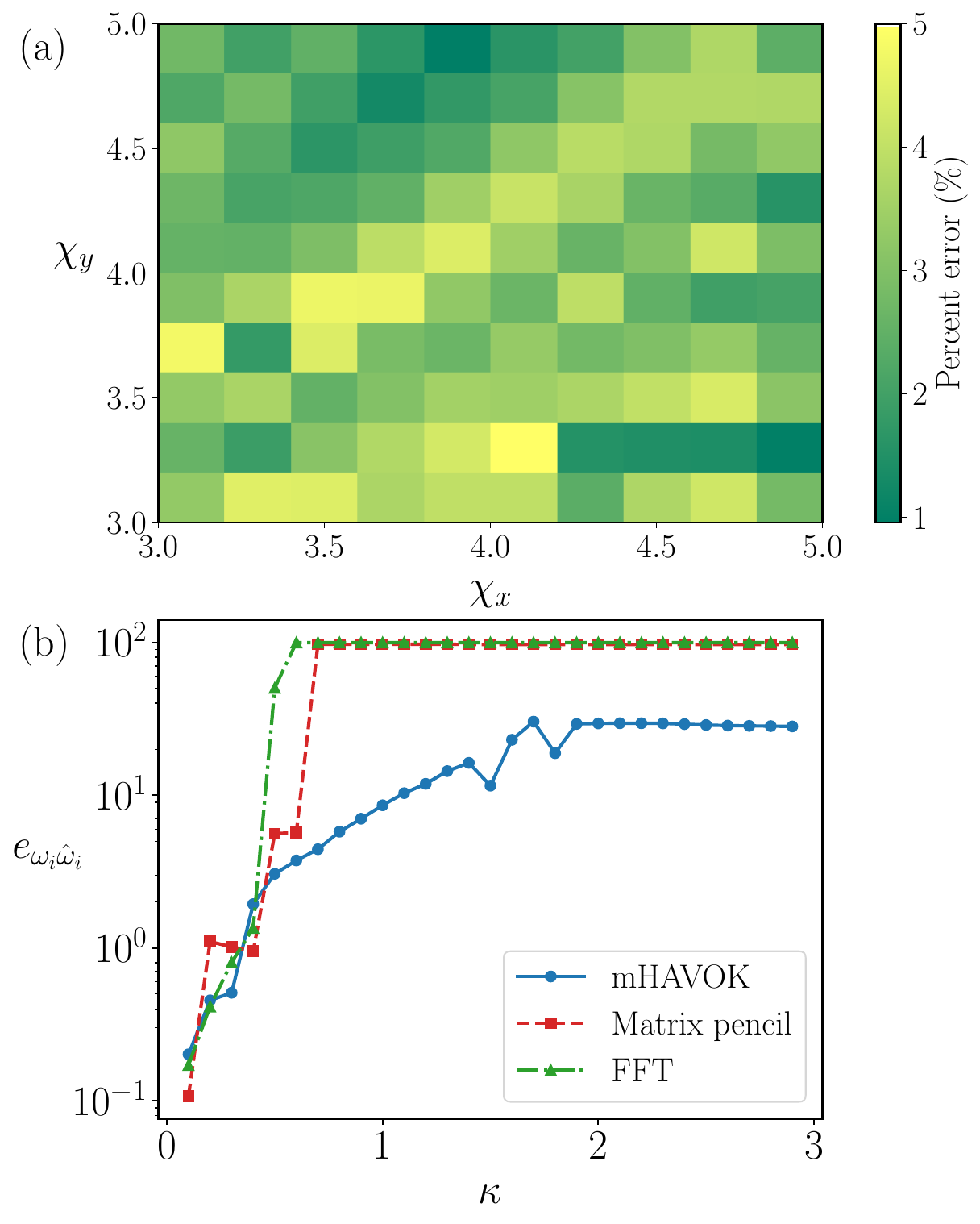}
    \caption{(a) Error comparison for a variation of Kerr nonlinear strengths. The percent error was below 5\%, even for strong nonlinearities. To achieve this, the optimal rank $r_{opt}$ was calculated and considered for each simulation. (b) Percent error comparison among three different methods for determining the oscillation frequencies $\hat{\omega}_i$ of the 2D QHO. Kerr nonlinearities were included, with strength parameters set to $\chi_x=2$ and $\chi_y=3$, respectively.}
    \label{fig:error_comparison}
\end{figure}

To observe the evolution of percent error for various Kerr strengths, the simulation was performed over a combination of values $(\chi_x=2+0.3n, \chi_y=3+0.2n)$, where $n\in\mathbb{N}$ ranged from 0 to 10. The optimal cutoff rank was calculated for each pair of strengths. Figure~\ref{fig:error_comparison} (a) shows that the percent error remains below $5\%$ across all cases, even when both $\chi_x$ and $\chi_y$ are relatively high (greater than $3.0$). Interestingly, the error in recovering the oscillation frequencies in the presence of strong nonlinearities remained below $5\%$. Additionally, as the damping rate increased, \mH outperformed the FFT and the matrix pencil, as Fig.~\ref{fig:error_comparison}(b) exhibits. In general, the error tended to increase when $\omega_y+\chi_x \approx \omega_x$, as closely spaced frequencies needed to be resolved. However, this increase remained relatively low, with the maximum being $4.96\%$ for the case when $\chi_x=4.1$ and $\chi_y=3.2$.

As an example, for the case where both $\chi_x=\chi_y=5$, i.e., strong nonlinearities, the average percent error for Kerr strength parameters was $e_{\chi_i\hat{\chi_i}}=1.3\%$, while $e_{\omega_i\hat{\omega}_i}=1.5\%$. Figures~\ref{fig:chi1}\textcolor{blue}{(b)} and~\ref{fig:chi1}\textcolor{blue}{(d)} show the reconstructed phase space for the above configuration, where now the trajectories form lobes, each corresponding to an instantaneous variation of the phase due to the amplitude-dependent term $\chi_i\expval{n_i(t)}$ discussed earlier.

\begin{figure}[t]
    \centering
    \includegraphics[width=\linewidth]{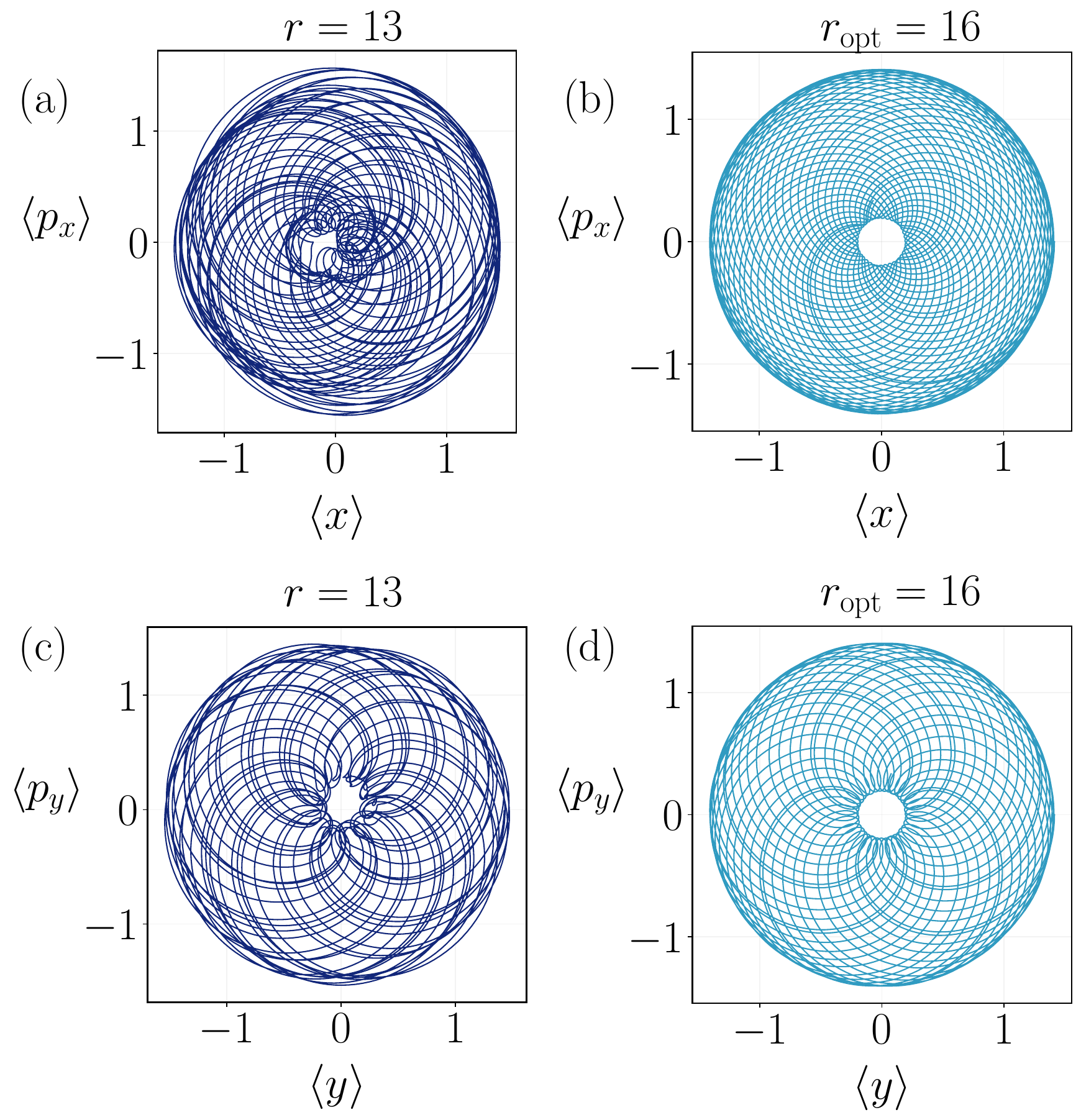}
    \caption{(a-d) Phase-space reconstruction of a closed system considering strong Kerr nonlinearities ($\chi_x=\chi_y=5$). The dark blue trajectories show the reconstruction using a cutoff rank close to (but not exactly at) the optimal value, while the light blue plots show the reconstruction using the optimal cutoff rank.}
    \label{fig:chix5_chiy5_closed}
\end{figure}

As a final comparison, we examined whether the algorithm could accurately reconstruct the oscillator shown in Figs.~\ref{fig:chi1}\textcolor{blue}{(b)} and~\ref{fig:chi1}\textcolor{blue}{(d)}, but under a closed system. Since \mH had successfully recovered the quantities of interest for the open system exhibiting strong Kerr nonlinearities, one might expect that reconstructing a closed setup would be straightforward. However, while the reconstruction was successful, as shown in Figs. \ref{fig:chix5_chiy5_closed}\textcolor{blue}{(b)} and \ref{fig:chix5_chiy5_closed}\textcolor{blue}{(d)}, the method proved to be highly sensitive to the cutoff rank $r$. This contrasts with previous cases, where choosing a rank close to the optimal one generally produced similar results. This is not generally the case, as illustrated in Figs.~\ref{fig:chix5_chiy5_closed}\textcolor{blue}{(a)} and~\ref{fig:chix5_chiy5_closed}\textcolor{blue}{(c)}. 

\subsubsection{Retrieving cross-coupling Kerr terms}

We also simulated several cross-term couplings $\chi_{xy}$, while keeping $\chi_x=0$ and $\chi_y=0$. 
In this case, the imaginary part of the eigenvalues of $\mathbf{A}$ is given by $\hat{\Omega}_i = \omega_i+\hat{\chi}_{ij}\expval{n_j}$, which is consistent with Eqs.~\eqref{eq:effective_frequencies_x} and~\eqref{eq:effective_frequencies_y}. After retrieving the oscillation frequencies $\hat{\omega}_i$, we determined the value of $\hat{\chi}_{xy}$ using the relationship mentioned above. To achieve this, we tracked the expected value of the number operator for both modes throughout the simulation using QuTiP.
After applying this procedure to various values of \(\chi_{xy}\), we summarize the corresponding results in Table~\ref{tab:cross_chi_values}. Generally, the method struggles to provide an accurate approximation for moderate and strong cross-coupling terms. This limitation arises because the forced linear model of \mH does not accurately reconstruct the evolution of the number operator. As a result, the oscillation frequencies obtained are only approximated based on the value of the number operator in the steady state.

\begin{table}[t]
    \caption{Predicted versus actual values of multiple cross-coupling Kerr nonlinearities.}
    \centering
    \rule{\linewidth}{0.5pt}
    \begin{tabular}{c c}
        Real value $\chi_{xy}$ & \hspace{2em} Predicted value $\hat{\chi}_{xy}$ \\
    \end{tabular}
    \vspace{0.3em}
    \rule{\linewidth}{0.4pt}
    \begin{tabular}{c c}
        0.05 & \hspace{8em} 0.07 \\
        0.10 & \hspace{8em} 0.09 \\
        0.15 & \hspace{8em} 0.18 \\
        0.20 & \hspace{8em} 0.19 \\
        0.25 & \hspace{8em} 0.15 \\
    \end{tabular}
    \rule{\linewidth}{0.5pt}
    \label{tab:cross_chi_values}
\end{table}

\subsection{Jaynes-Cummings interaction} \label{subsec:JC_results}

The Jaynes-Cummings model describes the interaction between a two-level atom, or qubit, and a single mode of the electromagnetic field \cite{jaynes_cummings}. For the following analysis, we assume a coupling on the $x$-mode of the electromagnetic field. In this situation, the qubit can exist in either the ground state $\ket{g}$ with energy $E_g$ or in the excited state $\ket{e}$ with energy $E_e$. 

The Hamiltonian of the system is $H=H_0 + H_{JC}$, where the Jaynes-Cummings Hamiltonian reads as
\begin{align}
    H_{JC}&=H_\text{atom} + H_\text{interaction},\label{eq:jaynes_cummingsHamiltonian}\\
    &=\frac{\Omega}{2}\sigma_z+g\hbar(a_x\sigma_++a_x^\dagger\sigma_-),\nonumber
\end{align}
where $\Omega\equiv E_e-E_g$, $g$ denotes the coupling strength, $\sigma_z$ is the third Pauli matrix, and $(\sigma_+,\sigma_-)$ represent, respectively, the raising and lowering qubit operators. Additionally, the qubit can spontaneously emit a photon and end up in the ground state. This procedure is known as relaxation and is modeled through a jump operator of the form \cite{jaynescummings_book}
\begin{align}
    L^\text{relaxation}=\sqrt{\gamma}\sigma_-,
\end{align}
where $\gamma$ is the relaxation rate. Similarly, the process by which the qubit's coherence decays is known as dephasing. If $\gamma_\phi$ denotes the dephasing rate, then this process is represented by the jump operator \cite{jaynescummings_book}
\begin{align}
    L^\text{dephasing}=\sqrt{\gamma_\phi}\sigma_z.
\end{align}

In this case, the equation of motion for the annihilation operator is coupled with the qubit lowering operator $\sigma_-$. Assuming the atom remains mostly in the ground state allows us to approximate $\sigma_z\approx 1$, leading to the linear set of equations
\begin{equation}
    \begin{pmatrix}
        \dot{a}\\ \dot{\sigma}_-
    \end{pmatrix}=-i\begin{pmatrix}
        \omega_x & g \\
        -g & \omega_q 
    \end{pmatrix}\begin{pmatrix}
        {a}\\ {\sigma}_-
    \end{pmatrix}.
\end{equation}

The eigenvalues of this matrix yield the dressed frequencies that describe the oscillations of the combined qubit-cavity system \cite{jaynes_cummings}. Mathematically, 
\begin{align}
    \omega_\pm=\frac{1}{2}\left[ (\omega_q+\omega_x) \pm \sqrt{4g^2+\Delta^2}\right],\label{eq:jaynescummings_dressedfrequencies}
\end{align}
with $\omega_q$ being the qubit frequency and $\Delta\equiv \omega_q-\omega_x$ representing the detuning, i.e., how far apart both oscillators are in frequency \cite{jaynescummings_book}.

Along with the simulation parameters presented in Table~\ref{tab:simulation_parameters}, we simulated the Jaynes-Cummings interaction with the initial state of the qubit $\ket{g}$ and an oscillation frequency $\omega_q=2\pi$. The coupling strength was set at $g=0.15$, while the relaxation and dephasing rates were, respectively, $\gamma=0.1$ and $\gamma_\phi=0.01$. For this case, \mH was provided with the observables
\begin{align}
    \mathbf{x}'=\begin{bmatrix}
    \expval{x(t)} \\ \expval{p_x(t)} \\
        \expval{y(t)} \\ \expval{p_y(t)} \\
    \expval{\Re(\sigma_+)(t)} \\ \expval{\Im(\sigma_+)(t)}
    \end{bmatrix}.\label{eq:mhavok_newinput}
\end{align}

\begin{figure}[t]
    \centering
    \includegraphics[width=\linewidth]{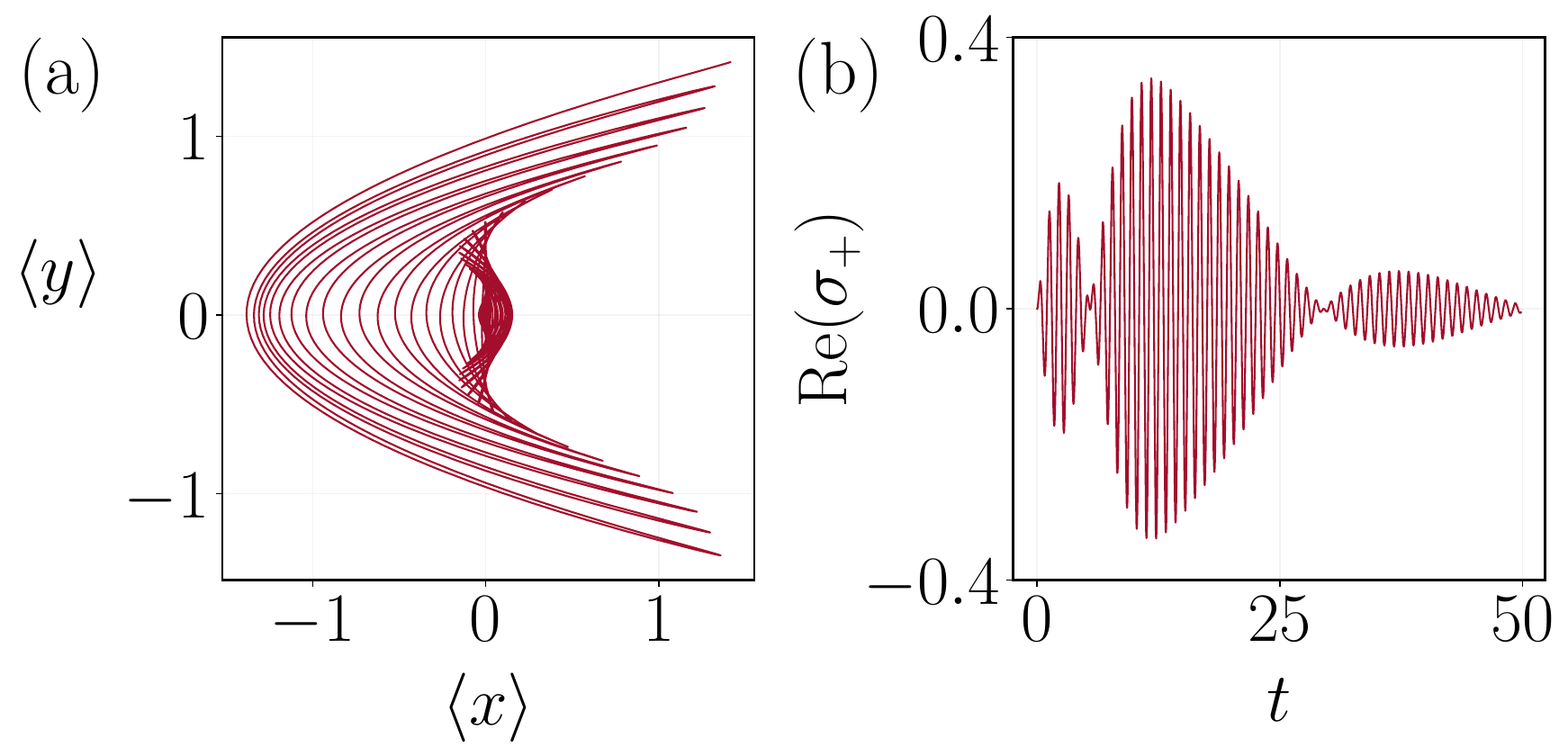}
    \caption{(a) Phase-space reconstruction under a Jaynes-Cummings interaction. (b) Retrieved evolution of $\expval{\sigma_+(t)}$ via \mH 's forced linear model.}
    \label{fig:jaynescummings_result}
\end{figure}

Figure~\ref{fig:jaynescummings_result}\textcolor{blue}{(a)} shows how the forced linear model successfully reconstructs the parabolic Lissajous curves with decaying amplitudes. In contrast to Fig.~\ref{fig:firstQHO_simulation}, this parabola does not collapse; instead, it changes orientation due to the interaction between the qubit and the field. Furthermore, the evolution of the raising qubit operator is also recovered and visualized in Fig.~\ref{fig:jaynescummings_result}\textcolor{blue}{(b)}. Here, coherent Rabi oscillations are prominent at early times, whereas the long-term behavior indicates decay, which can be attributed to the gradual loss of energy and coherence through the jump operators. The optimal cutoff rank was determined to be $r=20$, with all identified components classified as linear.

Moreover, the imaginary part of the eigenvalues of $\mathbf{A}$ enables the retrieval of the dressed frequencies defined in Eq.~\eqref{eq:jaynescummings_dressedfrequencies}. When solving for $g=0.15$ in this equation, a retrieved value of $\hat{g}=0.13$ was obtained, resulting in a percent error of $e_{g\hat{g}}=13.33\%$. These results were sensitive to moderate-to-strong dephasing and relaxation rates.
However, they improved substantially when the qubit frequency was not a multiple or exactly equal to $\omega_x$. For example, choosing $\omega_q=\pi/e$ resulted in an estimate of $\hat{g}=0.16$, which has a percent error $e_{g\hat{g}}=6.67\%$, half of that obtained when $\omega_q=2\pi$. This occurs because when the two frequencies coincide, the system is in resonance, making it more difficult to reliably identify the coupling strength. 

Varying over multiple values of $g$, the method accurately recovered only weak and moderate coupling strengths. For example, for $g=0.25$, the method recovered $\hat{g}=0.28$, yielding a percent error $e_{g\hat{g}}=12.00\%$. However, when the coupling strength was doubled to $g=0.50$, the method yielded a percent error of $26\%$. Therefore, the algorithm provides a reasonable estimate of the coupling parameter when the qubit frequency is not close to the field's oscillation frequency, and the coupling strength remains in the weak-to-moderate regime.

\subsection{Parametric frequency modulation} \label{sec:results_time_dependent_Hamiltonian}

The Hamiltonians analyzed so far have been time-independent. In this section, we demonstrate that \mH accurately recovers parameters in open systems with time-dependent Hamiltonians.
Let us consider the Hamiltonian
\begin{equation}
    H = H_0 + H_f(t),
    \label{Hft}
\end{equation}
where the 2D QHO Hamiltonian $H_0$ [Eq.~\eqref{eq:hamiltonian_field}] is modified by a time-dependent Hamiltonian with a modulation frequency on the $x$-mode of the form
\begin{equation}
    H_f(t)=\delta \cos(w_ft)a_x^\dagger a_x, \label{eq:timedependentHamiltonian1}    
\end{equation}
where $\delta$ denotes the modulated amplitude and $\omega_f$ the modulated frequency.

For the Hamiltonian in Eq.~\eqref{Hft}, the ladder operator and, hence, the quadratures evolve as 
{\small
\begin{equation}
    \langle a_x(t) \rangle = \langle a_x(0) \rangle 
    e^{-\kappa t/2} \sum_{m=-\infty}^\infty J_m\!\left( \frac{\delta}{\omega_f} \right) 
    e^{-i(\omega_x+m\omega_f)t}.
    \label{eq:time_dependentHamiltonian_ev}
\end{equation}
}

Equation~\eqref{eq:time_dependentHamiltonian_ev} exhibits sidebands at $\omega_x+m\omega_f$ weighted by the Bessel coefficients $J_m\left( \delta/\omega_f\right)$. In the presence of Kerr and cross-Kerr terms, these sidebands are shifted by the effective frequencies $\Omega_i^\text{eff}$, combining the effects of frequency and intensity-dependent phase rotations.

\begin{figure}[t]
    \centering
    \includegraphics[width=\linewidth]{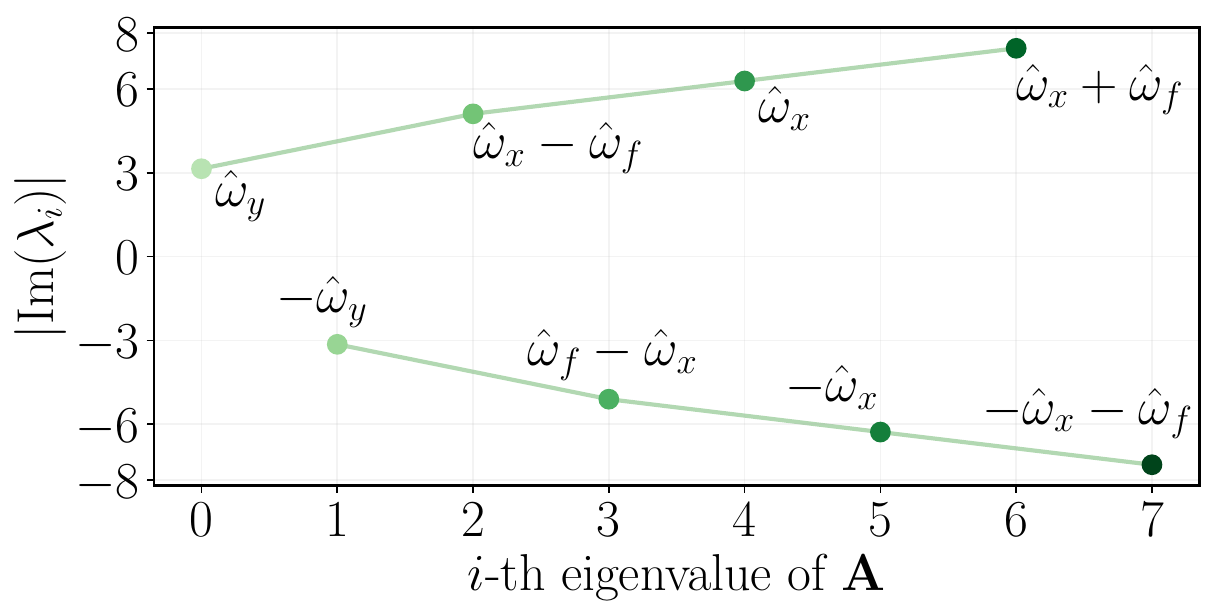}
    \caption{Evolution of the imaginary part of the eigenvalues of $\mathbf{A}$, which are denoted as $\lambda_i$. Complex-conjugate pairs are obtained, each representing a linear combination of the oscillation frequencies $\omega_x$, $\omega_y$, and $\omega_f$.}
    \label{fig:time_dependentHamiltonian}
\end{figure}

By simulating with a modulated amplitude $\delta=4$ and frequency $\omega_f=\pi/e$, we verified that the algorithm successfully recovered $\hat{\omega}_f$ within $1.74\%$ of their actual value. Figure~\ref{fig:time_dependentHamiltonian} shows how the imaginary part of the spectrum grows and recovers all oscillation frequencies, even when time-dependent Hamiltonians are present. These can be identified as the expected sidebands presented in Eq.~\eqref{eq:time_dependentHamiltonian_ev}. Moreover, as detailed in Sect.~\ref{sec:connection_Koopman_mHAVOK}, their appearance can be interpreted as eigenvalues of the Koopman generator restricted to the space of quantum observables, and whose realization is given by $\mathbf{A}$. This allows us to approximate a Koopman-invariant subspace from the observables, providing a well-behaved linear evolution of the dynamics, even in the presence of time-dependent parameters. To achieve this, all relevant frequencies must be captured, which is indeed the case.

As a limitation, the method had difficulty retrieving $\omega_f$ for large modulation coefficients $(\delta>3)$. For instance, when the modulation amplitude was increased to $\delta=10$, the spectrum exhibited no recognizable pattern. One of the retrieved frequencies was identified as $\hat{\omega}_y$ with an error of 0.06\%. However, the preceding and following frequencies could not be expressed as a linear combination of $\hat{\omega}_x$ or $\hat{\omega}_f$. 
As $\delta$ increases, it brings in contributions from more Bessel functions in Eq.~\eqref{eq:time_dependentHamiltonian_ev}, leading to the appearance of multiple neighboring and closely spaced sidebands. 
Therefore, the algorithm cannot isolate and retrieve the modulated frequency $\hat{\omega}_f$. However, the forced linear model can still reconstruct the system dynamics, primarily because the additional nonlinear dimensions captured in $\mathbf{B}$ are expressed in terms of multiple Bessel functions. Despite this capability, the model does not accurately retrieve the modulated frequency.

\section{Conclusions} \label{sec:conclusions} 

We demonstrated that the \mH method provides an efficient spectral framework for recovering the Hamiltonian parameters of open quantum systems directly from first-moment observables. By connecting the spectrum of the Koopman generator with the eigenvalues of the \mH matrix $\mathbf{A}$, we established a formal bridge between data-driven linear models and the underlying physical dynamics. This connection enables the discrete spectrum of $\mathbf{A}$ to serve as a reliable estimator of the relevant parameters of the system without requiring an analytic solution of the corresponding master equation.

By constructing a forced linear model for the expected value of first-moment observables, spectral analysis accurately recovers oscillation frequencies, damping rates, and Kerr nonlinearities, even in the presence of time-dependent Hamiltonians. A reasonable, though not ideal, estimation is achieved for recovering the coupling strengths of a Jaynes-Cummings interaction. Most recovered parameters differ from their actual values by less than 5\%, while one of the Jaynes-Cummings coupling strengths shows a maximum deviation of 15\%. In these cases, the \mH framework identified more nonlinear components, thereby increasing its influence on the system dynamics. Since spectral analysis is performed on the matrix that effectively captures the evolution of the linear components, a slight decrease in parameter retrieval accuracy is expected. Additionally, the matrix describing the nonlinear components is generally rectangular, which precludes spectral analysis.

An advantage of the \mH algorithm is its ability to recover strong damping rates, even in the presence of nonlinearities. This value was increased by almost an order of magnitude and was still accurately recovered, at least in the absence of Kerr nonlinearities or Jaynes-Cummings interactions. Furthermore, the method was compared with the FFT and matrix-pencil techniques for recovering oscillation frequencies and damping rates. It outperformed both techniques in strongly damped cases and yielded results equivalent to those of the other methods in other regimes. 

Finally, we remark that the method relies on Koopman operator theory to reconstruct the system dynamics. One of the dynamical matrices represents a finite-dimensional realization of the Koopman operator on a Koopman-invariant subspace spanned by the expected values of the observables. To the best of the authors' knowledge, this theoretical connection has not been previously reported. Therefore, such results motivate further exploration of Koopman theory as a practical framework for characterizing quantum dynamical systems.

\section*{Acknowledgments}

We acknowledge fruitful discussions with Dr. Oliver Probst's research and running group. 

\subsection*{Data availability statement}

The data that support the findings of this article are not publicly available. The data are available from the authors upon reasonable request.



\appendix

\section{Overview of the \mH method}\label{sec:appendixA}

The HAVOK algorithm was originally introduced as a method for studying classical nonlinear dynamical systems via forced linear models~\cite{Brunton_Havok_Nature}. The methodology was mainly applied to attractors and used Takens' embedding theorem, which guarantees a diffeomorphic attractor from delay embeddings of a  K-observable \cite{Takens_Embedding_Theorem}. Although HAVOK successfully reconstructs several dynamical systems, multiple limitations have been identified \cite{havok_limitations_1, havok_limitations3}: first, only one function of the system's state variables could be provided as input. Second, out of the $r$ main dynamical modes identified in the schema, it was arbitrarily assumed that the system's forcing only came from the $r$-th mode. Third, no objective criterion for selecting such $r$ was provided.

The limitations mentioned above were recently addressed in a new methodology named \mH ~\cite{mHAVOK}. The proposed extension allowed the method to include multiple time measurements, included a framework for identifying linear and forcing terms among the main $r$ dynamical modes, and presented an objective criterion for selecting $r$. Such improvements remarkably improved the reconstructed dynamics, even in complex nonlinear systems. Here we present an explanation of how \mH works; for a detailed discussion, we refer the reader to Ref.~\cite{mHAVOK}.

\subsection{Review of the \mH algorithm.}

Consider the evolution of the expected values of  K-observables of a (possibly nonlinear) dynamical system. In this work, we study the expected values of $n$ quantum observables, each represented by a $N$-dimensional vector and stacked in a vector $\mathbf{g}_k$,
\begin{align}
    \mathbf{g}_k:=
        \begin{bmatrix}
            \expval{\hat{O}_0(t_k)}\\
            \vdots\\
            \expval{\hat{O}_{n-1}(t_k)}
        \end{bmatrix},
        \label{eq:observables}
\end{align}
where $k\in\mathbb{N}$ ranges from $0$ to $N-1$. The method performs a delay embedding on $\mathbf{g}_k$ as follows:
\begin{equation}
    \mathbf{h}_k:=\left[\mathbf{g}_{k},\mathbf{g}_{k+\tau_s},\dots,\mathbf{g}_{k+(m-1)\tau_s}\right]^T,
    \label{deylemap}
\end{equation}
where $m$ is the embedding dimension and $\tau_s=1$ is known as the delay time. All $\mathbf{h}_k$ are stored in a block Hankel matrix,
\begin{align}
    \mathbf{H} = 
    \begin{bmatrix}
        \mathbf{h}_0 \quad  \mathbf{h}_1 \quad\cdots & \mathbf{h}_{N-m}
    \end{bmatrix}
\end{align}

Then, a Singular Value Decomposition (SVD) is performed on $\mathbf{H}$,
\begin{align}
    \label{SVD}
    \mathbf{H} = \mathbf{U}\mathbf{S}\mathbf{\mathcal{V}}^T,
\end{align}
where the first $r$ columns of $\mathbf{\mathcal{V}}^T$ are retained; let $\mathbf{V}$ denote this new matrix and $r$ be referred to as the cutoff rank. With this considered, a first regression procedure is performed on $\mathbf{V}$ in order to obtain $\mathbf{C}$, which is defined as the matrix that minimizes
\begin{align}
    \min_{\mathbf{C}} ||\mathbf{\dot{V}}-\mathbf{V}\mathbf{C}^T||_F, \label{eq:minimization} 
\end{align}
with $||\cdot||_F$ denoting the Frobenius norm and $\dot{\mathbf{V}}$ the temporal derivative of $\mathbf{V}$. Afterward, the coefficient of determination between each of the columns of $\dot{\mathbf{V}}$ and $\mathbf{V}\mathbf{C}$ is performed, 
\begin{align}
    \label{r_squared}
    R_i^2=1-\frac{||\mathbf{\dot{v}}_i-\mathbf{V}\mathbf{c}_i||^2}{||\mathbf{\dot{v}}_i-\bar{\dot{{v}}}_i\cdot\mathbf{1}||^2},
\end{align}
where $(\mathbf{c}_i, \mathbf{\dot{v}}_i)$ are, respectively, the \textit{i}-th columns of $\mathbf{C}$ and $\mathbf{\dot{V}}$, $||\cdot||$ denotes the Euclidean norm in ${L}^2$, ${{\bar{\dot{v}}}_i}$ is the temporal mean of $\mathbf{\dot{v}}_i$ and $\mathbf{1}$ is a vector of ones. Each of the $r$ columns of $\mathbf{V}$ is classified as linear or nonlinear according to their $R^2$ value:
\begin{align}
    \mathcal{I}_c&:=\{i:R_i^2\ge\tau\}, \label{eq:user_threshold}\\
    \mathcal{I}_f&:=\{1,\dots,r\} \backslash \mathcal{I}_c,\label{eq:user_threshold2}
\end{align}
with $\tau$ being a user-defined threshold and ($\mathcal{I}_c$, $\mathcal{I}_f$) representing, respectively, the index sets of linear and nonlinear dimensions. If $\mathbf{v}_i$ is the \textit{i}-th column of $\mathbf{V}$, then such components are separated into two different matrices:
\begin{align}
    \mathbf{V}_c := [\mathbf{v}_i]_{i\in\mathcal{I}_c}, \quad \mathbf{V}_f:=[\mathbf{v}_i]_{i\in\mathcal{I}_f}.
\end{align}
Next, a second regression problem is defined to obtain two new matrices; each models, respectively, the evolution of the linear ($\mathbf{A}$) and nonlinear ($\mathbf{B}$) components. Mathematically,
\begin{align}
    \label{eq:second_reg}
    \min_{\mathbf{A},\mathbf{B}} ||\mathbf{\dot{V}}_c-\mathbf{V}_c \mathbf{A}^T - \mathbf{V}_f \textbf{B}^T||_F,
\end{align}
where $\mathbf{A}\in \mathbb{C}^{|\mathcal{I}_c|\times|\mathcal{I}_c|}$ and $\mathbf{B}\in \mathbb{C}^{|\mathcal{I}_c|\times|\mathcal{I}_f|}$ and  $|\cdot|$ denoting the cardinality of the sets. Subsequently, the following forced linear system is numerically solved:
\begin{align}
    \label{eq:linear_model}
    \dot{\mathbf{x}}&=\mathbf{A} \mathbf{x} + \mathbf{B}\mathbf{u},
\end{align}
with $\mathbf{u}:=\mathbf{V}_f$ and $\mathbf{x}$ representing the simulated state vector. The initial condition $\mathbf{x}(0)$ is a column vector given by the first row of $\mathbf{V}_c$. Finally, the simulated block Hankel matrix $\mathbf{\hat{H}}$ is retrieved by reapplying the SVD factorization:
\begin{align}
    \label{inversetrans}
    \hat{\mathbf{H}}=\mathbf{U}_c\mathbf{\Sigma}_c \mathbf{x}^T,
\end{align}
where $\mathbf{U}_c$ and $\mathbf{\Sigma}_c$ represent the SVD matrices of the linear components. The reconstructed  K-observables defined in Eq.~\eqref{eq:observables} will be given by the first block of $\hat{\mathbf{H}}$. With this procedure, \mH provides a forced linear model that allows the quantities of interest to be obtained. 

\subsection{On the identification of the optimal cutoff rank}\label{subsec:optimal_cutoff_rank}

The optimal cutoff rank $r_\text{opt}$ is identified by calculating the condition number of the dynamical matrix $\mathbf{B}$ \cite{mHAVOK}, which is defined as the quotient between its highest and lowest singular values \cite{condition_number_matrix}. Mathematically,
\begin{align}
    \kappa(\mathbf{B}):=\frac{\sigma_\text{max}(\mathbf{B})}{\sigma_\text{min}(\mathbf{B})},
\end{align}
with $\sigma_\text{max}(\cdot)$ and $\sigma_\text{min}(\cdot)$ each representing the maximum and minimum singular values of $\mathbf{B}$. Physically, this value represents an optimal, non-redundant choice of dynamical modes from which the system can successfully construct a forced linear model of its dynamics. In general, exceeding such a value may lead to overfitting, while not reaching it may prevent crucial dynamical modes from being perceived, precluding a successful reconstruction from being obtained. The condition number, being an indicator of how sensible a matrix is to changes in its input \cite{condition_number_matrix}, successfully captures these phenomena, therefore being an objective criterion for the cutoff rank selection.

Considering each choice of $r$ has an associated $\mathbf{B}$, the condition number was calculated over a sweep of values $[r_\text{min}, r_\text{max}]$, with $r_\text{min}$ being the minimum number from which \mH identified at least one linear dimension and $r_\text{max}=30$ an arbitrary upper value. The optimal rank used throughout this work was the one with the highest condition number within the selected interval.

\subsection{On the selection of the embedding dimension and the number of employed 
\texorpdfstring{$K$}{K}-observables}

In general, the accuracy of the recovered parameters cannot be made arbitrarily good by increasing $m$ or by changing its value arbitrarily. For a given dataset, there exists an optimal embedding dimension beyond which the approximation does not improve and may even deteriorate due to overfitting. Thus, increasing the value of $m$ does not necessarily enhance the accuracy of the recovered parameters. Indeed, its optimal value has been a major focus of research in the delay-embeddings literature; we refer the reader to Ref.~\cite{optimal_embedding_dimension} for a comprehensive review of this topic.

On the other hand, increasing the number of observables $q$ tends to improve the approximation, as Ref.~\cite{deyle_sugihara} showed that, at least for classical delay embeddings, noise correlation is decreased by introducing more $K$-observables, notoriously improving the reconstruction of the system dynamics.

\subsection{Computational requirements demanded by \mH }\label{sec:computational_time}

Considering an input of $C$ vector-valued of  K-observables, each of length $N$, an embedding dimension $m$, and a cutoff rank $r$ in the SVD procedure, the computational cost of \mH is
{\small
\begin{multline}
    O\!\left[mC(N-m)\right]
    + O\!\left[mC(N-m)\,\min\!\bigl(mC,\; N-m\bigr)\right] \\
    {} + O\!\left[(N-m)r^{2} + r^{3}\right].
\end{multline}
}

The first term corresponds to the construction of the Block Hankel matrix, the second to its SVD factorization, and the third to the regression procedures and the state-space simulation of the forced linear model. In practice, $N>>mC$ and the SVD term dominates. Hence, the time complexity reduces to $O[(mC)^2N]$, i.e., it increases quadratically with the embedding dimension and linearly with the length of each vector-valued  K-observable.

\raggedbottom
\newpage


%

\end{document}